\documentclass[ preprintnumbers,A4paper,twocolumn,prb]{revtex4}
\usepackage{graphicx}

\newcommand{\ket}[1]{| #1 \rangle}

\begin{document}

{\Large \centerline  {\bf Wireless  computation in  }
\centerline { \bf self-assembled nanoscale arrays}}
\smallskip
\smallskip
\centerline {\bf Simon Charles Benjamin}
\centerline {\small Department of Materials, University of Oxford,}
\centerline {\small Parks Road, Oxford, OX1 3PH, UK.}
\smallskip
{\bf 
Ordered nanoarrays, i.e. regular patterns of quantum structures at the nanometre scale, have recently been synthesized in a range of systems. Here I explore a possible route to technological exploitation: assuming a simple form of bistability for the individual units, I study a form of array {\em computation} that is robust, efficient, programmable and highly defect tolerant. The nanoarray would need to be `wired' to conventional technologies {\em only} at its boundary; its internal dynamics are driven by intrinsic cell-cell interactions and global optical pulses addressing entire structure indiscriminately. Any self-assembled array would have a unique set of defects, therefore I employ an ab initio evolutionary process to subsume such flaws {\em without any need to determine their location or nature}. The approach succeeds for various forms of physical interaction within the array.}
\smallskip

There is an intense worldwide search, spanning both academic and commercial sectors, to find a realistic route toward computing with molecular scale structures\cite{ITRS}. 
One possibility which has deservedly received much attention, is that molecular-scale structures might be engineered to directly mimic the behaviour of today's transistors. However it seems entirely plausible that other architectures may be more `natural' for the exploiting the physics of the molecular scale. For example, IBM have presented a novel processing scheme which makes use of a molecular cascade phenomenon\cite{IBMthing}. Several other groups plan to exploit the fact that long molecules (such as nanotubes) can conduct, by employing ingenious architectures based on a dense grid of wires\cite{HP,lieber}. In this paper I focus on the exact opposite, or compliment, of a wired structure: at the smallest scale the device is formed from a regular array of {\em isolated} elements, which I refer to generically as {\em cells}. There is no flow of charge between cells; their communication is purely via fundamental physical interactions such as magnetic dipolar or electrostatic forces. 

Disordered nanoarrays, with random cell locations and sizes, have been observed in many systems. For example, layers of randomly distributed quantum dots can be routinely synthesized in III-V semiconductors. However, in recent years {\em ordering} has been successfully achieved in several classes of system. One family of techniques involves manually defining a template for cell nucleation via scanning-tunnelling microscopy\cite{kohmoto}, or nanoimprinting\cite{kamins}. There have also been recent advances in using dip-pen nanolithograph to direct nanoparticle location, e.g. by prior deposition of DNA sequences\cite{smallGingerMirkin}. These powerful techniques could create complex arrays, but may be too labour-intensive for commercial application. The alternative is the broad family  of pure self-assembled systems, where periodic patterns form without any `top-down' intervention (as will be described presently, even simple periodic patterns can implement general circuits). Several recently introduced techniques for self-assembly involve co-opting biological systems:  protein rings harvested from engineered bacteria can produce ordered arrays of sockets \cite{proteinRingForDots}; engineered viruses can capture nanoparticles and integrate into an array \cite{dotArrayAssembledByVirus}; DNA `tags' can be attached to nanoparticles\cite{Mirkin607, nanoLettersSeeman} which can then be attached to 2D periodic DNA  scaffolding \cite{nanoLettersSeeman, DNAcrystal, turberfield05}. Complex structures can arise through an interplay between the dimensions of host templates and guest molecules~\cite{beton}, and studies at the millimetre scale suggest that this phenomenon can be exploited in sophisticated ways~\cite{mmTemplating}. Engineering of the self-assembly process can yield structures with novel optical properties~\cite{templatedOpticalColumns}, and entities with multiple (and switchable) internal states~\cite{multiLevelMolecMem}. Ordering can also occur without any scaffolding: fullerenes can be functionalised to form an ordered superlattice~\cite{wattEtAl} while quantum dots can be prepared via colloidal chemical synthesis and subsequently deposited on a surface\cite{bronmann}, or may grow directly on the surface with\cite{Jin} or without \cite{Jian} prepatterning. Li {\em et al}\cite{Jian} have produced dots of strictly identical size, ordered into a hexagonal array with atomic precision (complimentary to the Si(111)-(7$\times$7) substrate). Moreover they have produced periodic arrays with more than one type of dot interspersed in a regular pattern. Recent work has shown that binary arrays can be created with a startlingly diverse range geometric patterns~\cite{natureTalapin}. 

I hope that the present paper will help to motivate further efforts to study and engineer the cell states and interactions within such structures. By making some plausible assumptions for these properties, I describe a novel mechanism by which nanoarrays can {\em compute}. The form of computation considered here is classical information processing, even though the information is represented by quantum states. This is to be distinguished from true quantum information processing (QIP), which involves maintaining and manipulating superposition and entanglement. The demanding nature of QIP means that large arrays will not achieve it without first achieving classical computation. At the end of this paper I will briefly consider extending the ideas to QIP. 

There is a substantial existing literature on the topic of finding novel mechanisms for nanoscale classical computation. A few characteristic examples include the work of Likharev and Korotkov\cite{Lik&Kor}, by Bandyopadhyay\cite{Bandy}, and by Johnson and myself\cite{benj2DCA}. Perhaps the most well known idea is that of Lent, Tougaw {\em et al}, which (in its original form) involves constructing an array in such a way that simply relaxing to the ground state corresponds executing a computation \cite{smithRevsLent}. Meanwhile, there is an extensive literature on the related mathematics of cellular automata (see eg Ref. [\onlinecite{tof}] for an introduction) and the study of evolutionary principles as an approach to circuit design is an active sub-field in computer science \cite{GPEM,ThompsonPaper,evolArithCircuit,evoFPGA}. I draw on several of these ideas in this paper - however, to my knowledge no previous author combines the following factors into a single scheme: 
\begin{enumerate}
\item \label{nice1} 
Elementary bistable quantum cells are ordered in a {\em simple regular array}, with no need for complex patterning and therefore being suitable for self assembly. However, the array can nevertheless be programmed for arbitrary computational tasks (logical circuits).
\item \label{nice2} All control is via global pulses that address the entire array indiscriminately (no need for local wiring/clocking of cells within the array).
\item \label{nice3} I employ dissipative, one-way switching process with a highly non-linear response curve - characteristics associated with the success of conventional micro-technology. However, (a)  there is minimal energy dissipation per gate, and (b) the majority of waste energy can exit optically rather than thermally (preventing overheating). 
\item \label{nice4}  There is an efficient means of finding functional behavior despite various physical array defects - even when such defects cannot be identified (again, vital for realistic self assembled systems).
\end{enumerate}

\begin{figure}[!b]
  \begin{center}
    \leavevmode
\resizebox{8.3 cm}{!}{\includegraphics{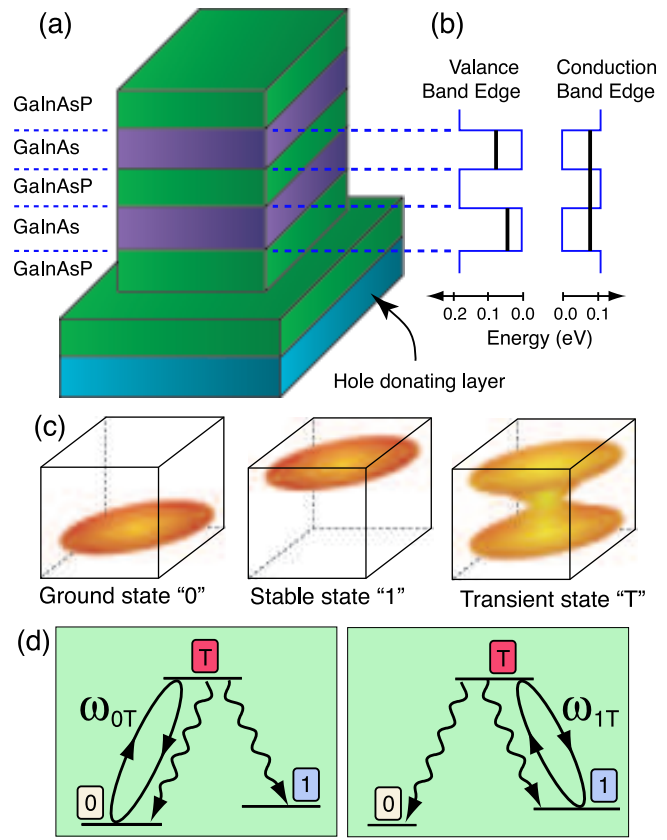}}
\end{center}
\caption{One example\cite{mahler} of the type of nanostructure that can serve as a cell. Essentially, there is a single excess hole in the structure which has one of two possible stable localisations, $\ket{0}$ and $\ket{1}$. (a) Physical structure. (b) Highest valence band hole states, and lowest conduction band electronic state. Bulk material band edge shown for context. (c) Schematic of the charge distribution in states $\ket{0}$, $\ket{1}$, and the optically excited transient state $\ket{T}$. (d) An isolated cell would be switched from state $\ket{0}$ to $\ket{1}$ via $\ket{T}$ by laser pumping at a frequency $\omega_{0,T}$. The reverse switch is achieved by frequency $\omega_{1,T}$.}
\label{figMahlerStructure}
\end{figure}

\begin{figure}[!b]
  \begin{center}
    \leavevmode
\resizebox{8 cm}{!}{\includegraphics{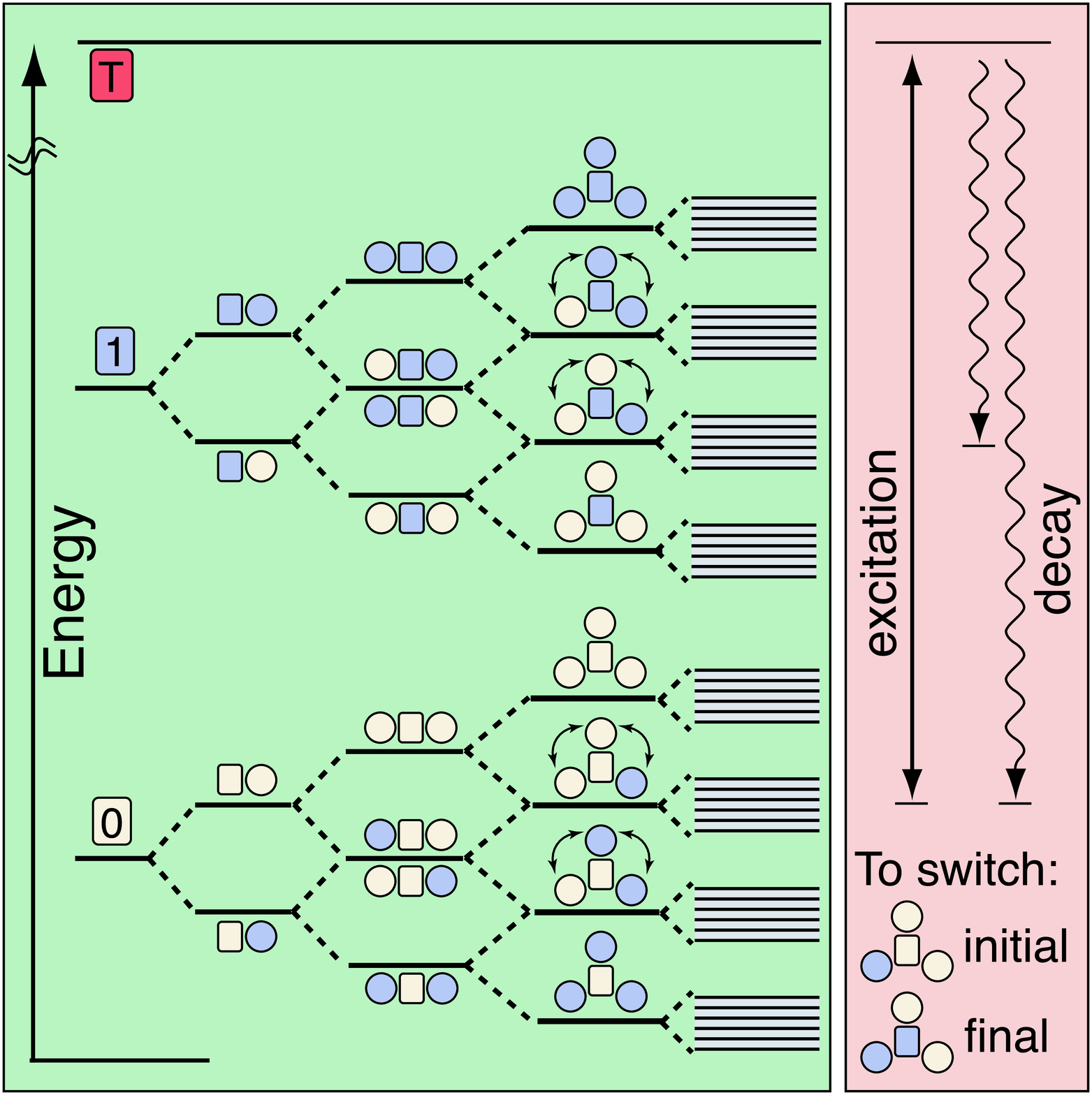}}
\end{center}
\caption{Left: Energy level structure for a single cell (depicted as a square), depending on the states of the neighbouring cells (circles). Each additional neighbour would result in a systematic shift common to all levels - this is omitted since we are concerned with the transition energies between levels. State $\ket{T}$ would be far above the lower states, as indicated by the broken $Y$ axis. The splittings can be regarded as shifts to the cell's basic switching frequencies $\omega_{0,T}$, $\omega_{1,T}$ {\em conditional} on the states of the nearest neighbours. The final column shows the discrete levels dissolving into bands due to the effect of non-nearest neighbour cells. These bands should be non-overlapping for reliable cell switching; given a specific form for the cell-cell interaction, this becomes a constraint on how perfectly ordered the array is (c.f. Fig. \ref{figComprehensive}). Right: optical switching can occur as in Fig. \ref{figMahlerStructure}, except that now we may selectively switch according to the states of the neighbouring cells. The neighbour's states themselves must be static, as discussed later. }
\label{figMahlerElevels}
\end{figure}

\begin{figure}[!t]
  \begin{center}
    \leavevmode
\resizebox{8 cm}{!}{\includegraphics{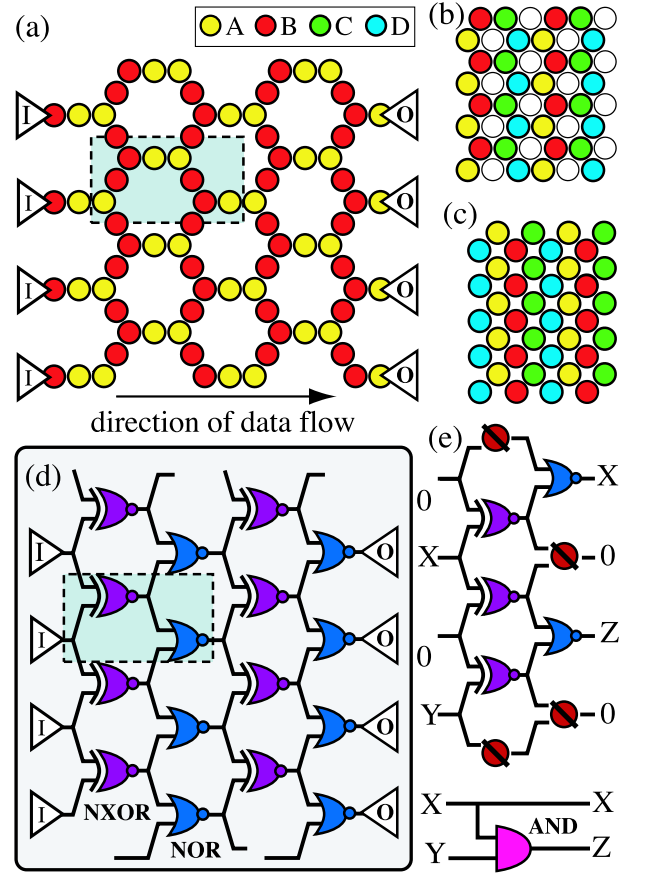}}
\end{center}
\caption{(a)-(c) Examples of the many periodic arrays that can support computation via the method described here. Data input/output occurs only at the sides, as denoted by triangles. With a suitable sequence of global pulses, binary states are driven through the array (left-to-right) in such a way that the array acts as a network of logical gates (d). The dotted square indicates corresponding areas of (a) and (d). Part (e) illustrates how an arbitrary function (here a simple AND circuit) can be programmed into the periodic gate array purely by allowing/disallowing specific data pathways - this routing can be achieved simply by pre-loading the array with a suitable set of states.}
\label{figConcept}
\end{figure}

The structure of this paper is as follows: first I describe the underlying theoretical model which I adopt for the cells and their interactions. In order to provide a physical example, I describe one suitable form of cell which has been previously discussed in the literature. I then describe the array dynamics which arise from the model in the limit of a certain hierarchy of energies. 
I explain that, in the ideal case of zero defects and pure nearest neighbour interactions,  a suitable sequence of global laser excitation pulses can exploit the natural dynamics to perform computation. (The explicit description of this pulse sequence, and its effect, is given in the Appendix.) I then report the results of an intensive numerical simulation, demonstrating that the same functionality can be achieved in realistic defective arrays. The approach involves a genetic algorithm - details are given in the Appendix. The Appendix also contains brief discussions of  the power and usefulness of an array computer (data processing versus data transport), along with remarks about the input/output interface, three dimensional arrays, and quantum information processing. These topics are placed in the Appendix in order that the main paper can give a compact overview of the research.

\bigskip
\bigskip
 
\noindent {\bf Summary of the underlying model} 

Each cell is taken to be a three state quantum system: states labeled $\ket{0}$ and $\ket{1}$ are stable, e.g. the ground state and a long lived excited state, and $\ket{T}$ is a rapidly decaying transient state which can be accessed by laser excitation. Real cells may have a more complex sprectrum, but provided that one can identify three such states then this model is appropriate. Given the broad range of array synthesis methods that have already been demonstrated, and the flexability of those methods, it is probable that such spectra can indeed be engineered. In order to provide a definite example, Figs. \ref{figMahlerStructure} and \ref{figMahlerElevels} depict a suitable cell that has previously received a rigorous theoretical analysis in the literature\cite{mahler}. However, this case is only one of many possibilities - it is equally appropriate to think in terms of a bistable molecule, for example. In any such three level system, the Hamiltonian of an isolated cell can be written $H_0= \sum_{p=0,1,T} E_p a_{p}^\dagger a_{p} $, and in the presence of a laser\cite{fn1} tuned to a frequency $\omega$ near a transition energy $\omega_{i,T}\equiv (E_T-E_i)/\hbar$,  one has $H^{\rm cell}= H_0+\hbar \omega (b^\dagger b + {1\over 2})+ \hbar R(a_{i}^\dagger a_{T}b^\dagger + b a_T^\dagger a_i)$. I take the coupling of the cell to its `heat bath' environment to be dominated by dissipative decay from the transient third state to either $\ket{0}$ or $\ket{1}$, i.e. I assume $\hbar \omega_{2,i}\gg k_bT$. An optical transition would therefore be suitable in a room temperature environment. The decay channel may be of any kind  (radiative, non-radiative, composite), although for efficient heat dissipation it is highly desirable if the decay is photonic.
The cell's density matrix $\rho$ is then governed by
$\dot{\rho} = -{i \over \hbar}[H^{\rm cell},\rho]+\dot{\rho}_{\rm incoh}$ where
$\dot{\rho}_{\rm incoh}=\sum_{p=0}^1 {d\over 2}([a^\dagger_p a_T,\rho a^\dagger_T a_p]+[a^\dagger_p a_T \rho, a^\dagger_T a_p])$.
It is established\cite{mahler}
that this form of master equation leads to one-way switching: if the cell is in state $\ket{0}$ when a pulse with frequency $\simeq\omega_{0,T}$ is applied (Fig. \ref{figMahlerStructure}d), then probability amplitude is switched to $\ket{1}$ exponentially quickly: $\rho_{1,1}\approx 1-\exp(-\lambda t)$. A pulse duration of $25/d$ is adequate for reliable switching within arrays of the size considered here. Moreover if the laser is adjusted to a frequency near $\omega_{1,T}$, then one can drive the reverse switch from state $\ket{1}$ to $\ket{0}$ (given that $\Delta\omega\equiv \omega_{1,T}-\omega_{0,T}$ is large on the scale of $R$, so that switching of the off resonance transition occurs negligably slowly).
I will assume that cell-cell interactions within our $N$-cell network are diagonal in the basis of the single-cell eigenstates, i.e. $H_{\rm int}={1\over 2}\sum^{1\cdots N}_{k\neq m}\sum_{p,q}^{0,1,T} K_{k,m}^{p,q} a_{k,p}^\dagger a_{m,q}^\dagger a_{k,p}  a_{m,q}$. Note that by no means all arrays will have such an interaction form: by making this assumption I am specializing to those that do \cite{fn2}. The effect of $H_{\rm int}$ is to shift all cell transition energies according to the states of neighbouring cells, thus giving us the basis for {\em conditional} switching (Fig. \ref{figMahlerElevels}). This is a dissipative, one-way switching process with a highly non-linear response curve -- the success of transistor technology is founded on the same combination, which is missing from many previous novel computing schemes \cite{smithRevsLent}. 
For simplicity I assume an isotopic dependence on cell separation, $K_{k,m}^{p,q}=f(|{\bf r}_k - {\bf r}_m|) K^{p,q}$. I will presently model three different forms for $f()$, varying from long to short range: $r^{-3}$, $r^{-6}$ and $\exp(-r^2)$. The contribution from the transient state $\ket{T}$ is assumed to be negligible ($K^{T,i}=K^{T,2}=0$). I make the Ising-type choice of $K^{p,q}=(-1)^{(p+q)}K$. The alternative form $(-1)^{(p+q+1)}$, i.e. swapping from anti-alignment to alignment, would be equally suitable. 

\begin{figure}[!h]
\centering
\resizebox{7.cm}{!}{\includegraphics{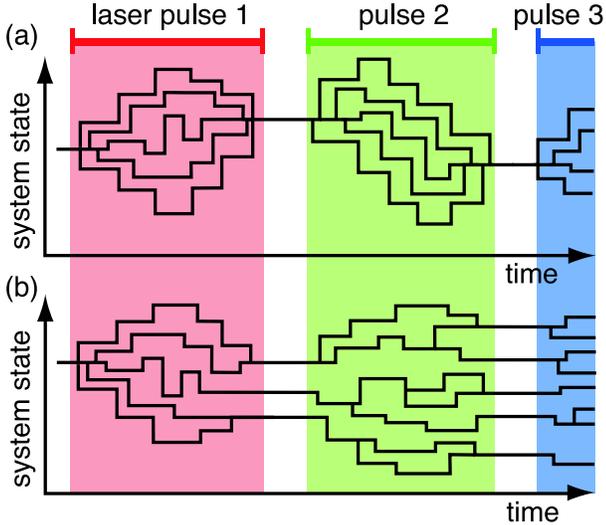}}
\caption{(a) Typically numerous cells within the array are eligible to be switched by a particular pulse. When such a pulse is applied globally to the entire array then those cells will switch at random times, and thus the system's state space trajectory diverges into numerous possible paths. However, because of the stable neighbourhood condition, the random switch order has no effect on eligibility and the paths converge into a well defined final state. (b) Without this stability condition, each global pulse could have several potential stable final states. Subsequent pulses would lead to further division, rapidly leading to a highly complex probabilistic mix of states.  }
\label{figDivergence}
\end{figure}

\bigskip

\noindent {\bf Array dynamics} 

Suppose that one has a distinct hierarchy of energies: laser band width $\delta<K<\Delta\omega\ll \omega_{i,T}$ 
 where symbols are defined above. Then the array system behaves as a set of {\em classical} units and is relatively straightforward to model numerically. It has the following properties: 
 
\begin{enumerate}
\setcounter{enumi}{0}
\item \label{array1} 
Cells can have their internal states `switched' (e.g. from $0$ to $1$)  by an externally applied optical pulse which targets the entire array indiscriminately. 
\item \label{array1} Each cell's susceptibility to a  given switching pulse is determined by the states of the immediately surrounding cells. 
\end{enumerate} 
Thus we have conditional state switching, which permits very complex patterns to develop in response to a sequence of pulses. The process is inherently {\em asynchronous} on the time scale of individual switching events: there is no way to control the precise moment a given cell switches during a pulse. To ensure a deterministic final state for the complete array, we must guarantee that when a given cell switches it state, other cells in the immediate neighbourhood remain static (see Fig.\ref{figDivergence}). Then the system effectively becomes {\em synchronous} on the time scale of the complete pulse. This is then a third array property which we must introduce:
\begin{enumerate}
\setcounter{enumi}{2}
\item \label{array3} 
While a given cell switches it state, other cells in the immediate neighbourhood remain static. 
\end{enumerate}
This is achieved by employing more than one `type' of cell. I will refer to the cell types by the letters $A$ and $B$, so that the transition energies are written $\omega^A_{1,T}$ etc. To ensure that a cell of type A is never switched by the same laser pulse that switches type B, we can require that the difference in their transition energies is much larger than $K$, i.e. $\omega^A_{i,T}-\omega^B_{j,T}\gg K$ for all $i,j=0,1$. 
It is then straightforward to derive array patterns that have this property: the patterns in Figure \ref{figConcept}(a)-(c) are examples. Array (a) has only two physical cell types, $A$ and $B$, but there are effectively four types: nodal cells have three neighbours, giving them different transition energies (although this proves to be less robust versus misalignment than four physical types - cf Fig. \ref{figComprehensive}b). Each array (a)-(c) is plausible experimental goal given the broad range of array geometries that has already been realized. Here I elect to focus on the sparse hexagonal geometry shown in Fig.  \ref{figConcept}a, which happens to be convenient for study\cite{whyCon}. Physically the distinct cell types may correspond to two different sizes of cell, or two different compositions, as in Ref. \onlinecite{Jian} where cells alternate between In and Mn. 

\bigskip
\bigskip

\noindent {\bf Idealized case.} 

Consider a {\em conceptually idealized} array: one that is defect free and has a strict nearest neighbour interaction. Then one may be able find a sequence of laser frequencies that has the effect of systematically driving states from left to right through the array. Two key sequences are required: one to simply load states into the array (to `program' it) and one to process data according to the program (illustrated in Fig.  \ref{figConcept}). Finding suitable sequences for a given array geometry is an interesting theoretical exercise; the Appendix specifies the sequences which I derived for the hexagonal array employed here. There I also explain that a  single array can be working on multiple sets of data simultaneously, i.e. it supports dense pipelining, leading to an excellent ratio of data bits to cells.

\begin{figure}[!t]
\centering
\resizebox{7.5cm}{!}{\includegraphics{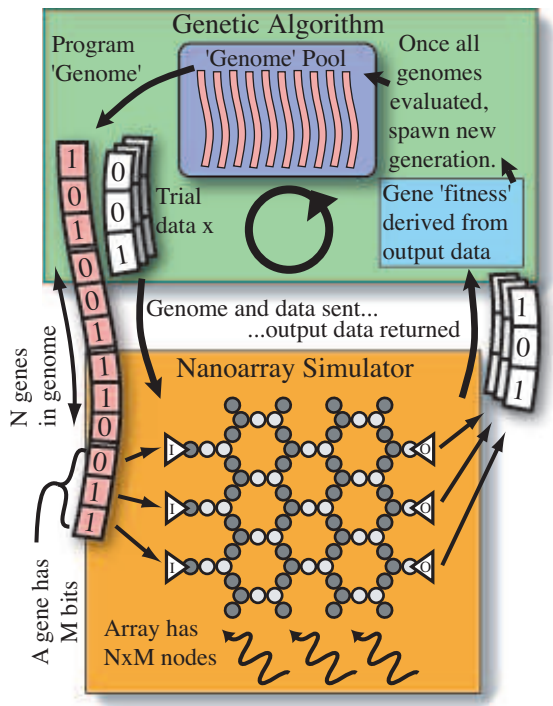}}
\caption{A numerical experiment has two independent parts: the nanoarray simulator (NS) and the genetic algorithm (GA). In a real experiment the NS would be replaced by the physical array and its surrounding apparatus, but the GA would retain its role. The NS begins by defining a unique new array, complete with randomly generated defects. The GA treats it as a `black box', whose internal workings are unknown, and proceeds to look for a way to make it perform a useful function $g(x)$. Initially the GA establishes a pool of  `genomes', each a random binary string. Each member in turn is evaluated: (1) The genome and a certain binary number $x$, are sent as input to the NS. (2) The NS loads the genome, gene-by-gene, onto the input cells while subjecting the array to the `programming' optical pulse sequence. Then it loads $x$, applies the `processing' sequence, measures the output and reports it to the GA. (3) Steps 1 \& 2 are repeated with different $x$. (4) The GA assigns the genome a fitness rating depending on how closely the outputs match $g(x)$. After all genomes are evaluated, the GA creates a new generation by random mutation (bit flipping). The fittest genomes produce multiple descendants, the least fit produce none. }
\label{figGenetic}
\end{figure}
\bigskip

\noindent {\bf Realistic arrays.} 

Real nanoarrays will typically have a cell-cell that is not strictly nearest-neighbor, but rather has a strength that falls as some smooth function of separation - how long range can the interaction be? Perfect periodicity over large regions is also unrealistic - even allowing for advanced fabrication techniques that might include `healing' of defects~\cite{selfHeal}, one must assume that there will be some characteristic defect rate. Can the same functionality established in the ideal case be achieved in an array that contains defects of an unknown type and number?  And can one succeed using data input/output only at the array edge? The use of evolutionary principles for circuit design constitutes an entire sub-field of computer science\cite{GPEM,ThompsonPaper,evolArithCircuit}. For example, research into field programmable gate arrays,\cite{evoFPGA}  a conventional microelectronic architecture that has similarities to our nanoarrays, has successfully exploited evolutionary principles. Therefore I attempt an analogous strategy, evaluating it via a series of intensive numerical experiments. 

\begin{figure*}[t]
\centering
\resizebox{16.cm}{!}{\includegraphics{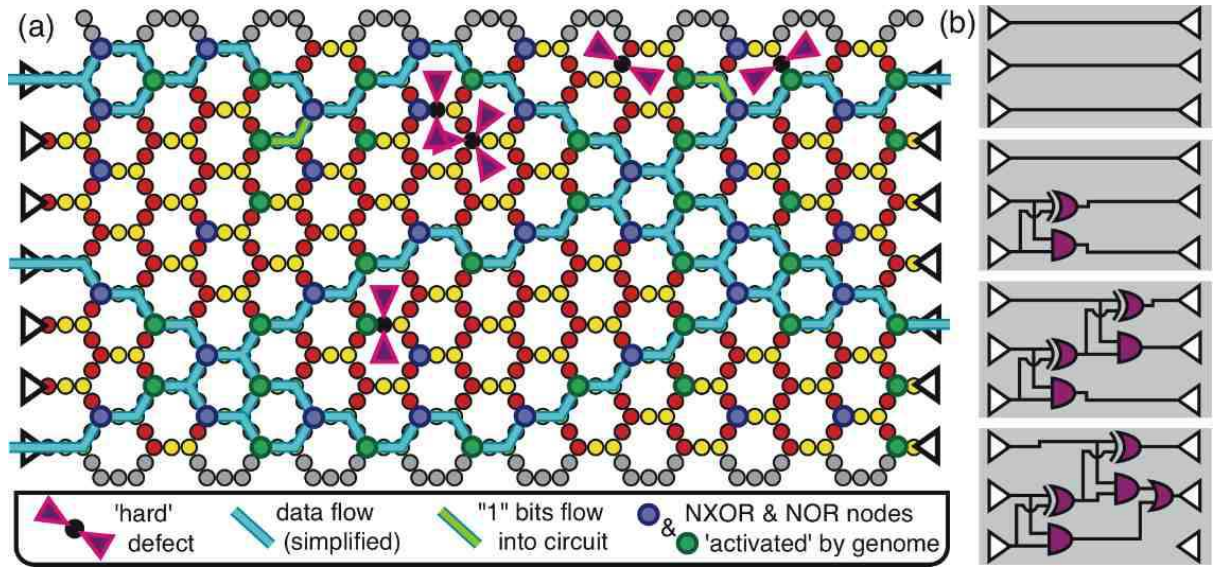}}
\caption{(a) Schematic showing the final state of the nanoarray in a particular experiment. The defective structure has been made to function as a sophisticated and useful circuit: a `full adder', the building block of binary arithmetic. Array defects include $5$ randomly located completely non-functional cells (`damage'), and random deviations in the positions of all cells (`misalignment') causing a $12\%$ variation in cell spacing. (b) A multi-stage evolutionary process was used: once a circuit is achieved perfectly, the goal switches to a more sophisticated target. In this way complex circuits can be developed without getting `stuck' in an evolutionary dead end.}
\label{figFullAdder}
\end{figure*}

\bigskip

 \begin{figure*}[t]
\centering
\resizebox{16cm}{!}{\includegraphics{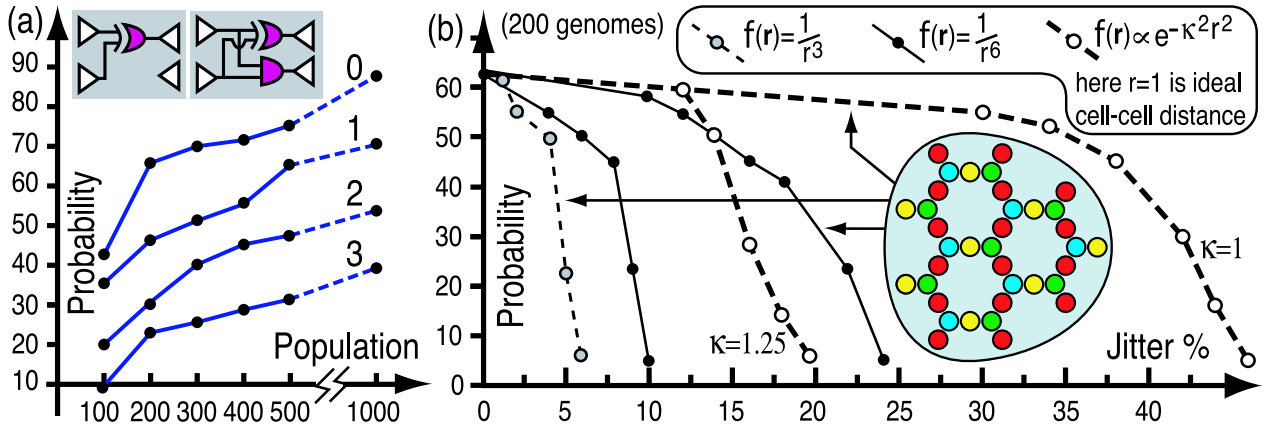}}
\caption{Graphs showing the probability of successfully implementing a `one bit half-adder' circuit, with the intermediate goal of a single {\small XOR} gate as show in the top-left inset. As explained in the text, the absolute probability of success depends on the length of time for which one runs the evolution - one must impose some arbitrary limit in order to make comparisons and expose trends. Here the constraint is that each phase must complete within 50 generations. In graph (a), successive lines show the impact of $0$, $1$, $2$ \& $3$ damage sites (i.e. non-functional cells). The points on the uppermost line would all lie at $100\%$ given infinite generations - it therefore provides a context for the other plots. The array contains 250 cells (half the size of Fig. \ref{figFullAdder}), thus $>1\%$ of such defects have been tolerated. Graph (b) exposes the effect of misalignment: random variations in all cell positions, leading to a range of cell separations, expressed as a percentage of the ideal separation. Successive lines correspond to various interaction ranges $f(\bf{r})$ and two array geometries: two-type, as in Fig. \ref{figConcept}b, and four-type as in the inset here.}
\label{figComprehensive}
\end{figure*}

In order to make this document self-contained for readers unfamiliar with these techniques, I include a complete description of the process I use. Figure \ref{figGenetic} shows the structure of each numerical experiment (for discussion of the interface at the sides of the array, see the Appendix). The overall procedure can be characterized by a loose analogy with biological evolution. The nanoarray simulator (NS) effectively determines the adult form of an animal given that animal's genome, and reports the behavior of the animal when it is provided with a certain sensory input $x$. The genetic algorithm (GA) calls for this process to be repeated for various different $x$, and then assigns a score to the genome depending purely on the reported behavior, according to some ideal target. 

I emphasize that the GA assigns a fitness score purely by analyzing the correlation between the recorded inputs and outputs. There is no attempt to infer the location or nature of defects, nor to make any other analysis of the internal physics of the array. Indeed the GA does not `know' anything about the NS - from the GA's perspective, the NS is simply a `black box'. The score assigned to each gnome is reduced if there is randomness - i.e. if repeated application of the same input yields differing output.  
The iterative process of evaluating the gnome pool, and generating a new pool, finally terminates when some genome exceeds a certain fitness score. This threshold corresponds to the nanoarray reliably producing the desired $g(x)$ for every input $x$. Figure \ref{figFullAdder} shows the result of one particular numerical experiment on a large defective array. Here a perfect genome was found in the 129$^{th}$ generation with a population of 1200, thus the total number of genomes tested was $154,800$. By comparison the total number of possible genomes is $2^{7\times16}\approx 10^{34}$. Array output exhibited substantial instability during the evolution, but the final genome implements a completely stable circuit.

In order to explore the effect of different numbers and types of defects, and different forms of cell-cell interaction, I performed an extensive series of numerical experiments. Because the GA process involves random mutations, even in an experiment on a defect-free array the process can sometimes get stuck
in an evolutionary dead end --
in which case it must be restarted form an earlier stage. This 
can even happen more than once, thus there is no time scale on 
which one can guarantee success, even for arrays where success is clearly
possible. Therefore I only permit the GA to run for a finite number of generations $N_G$ in reaching each phase target - if $N_G$ is exceeded the run is terminated and labelled a failure. Choosing $N_G=50$ gives any overall success probability high enough to expose the significant trends.
I could have run for twice as long, in which case the probability of failure
for defect free arrays would be halved (at least) - but since the purpose of the
graph is to reveal trends, this would not be fruitful.
Notice that in graph (b) the different forms of interaction converge with one another
in the limit of zero alignment error - in this limit suitably broad laser pulses can completely
`wash out' the beyond-nearest-neighbour splitting without ever stimulating a transition in an adjacent band. All five arrays then behave identically and perfectly - the success probability would be $100\%$ if I allowed unlimited generations. As 
successive experiments introduce some jitter, the spread of energy levels broadens beyond the band width of the applied pulse - and thus 
we get regional instability that the process must `learn' to avoid. As one would expect,
arrays with longer range interactions are more susceptible to this effect. However, arrays with more cell types (as shown in inset in Fig. \ref{figComprehensive}b) are seen to have superior tolerance of misalignment defects. By guaranteeing that, e.g., a given cell type only ever has an odd number of nearest neighbour cells, we are increasing the frequency space around each sub-level. 
Defects in cell transition energy would have a very similar effect to the positional defects analyzed here.

 \bigskip
 
\noindent {\bf Discussion and Conclusion.} 

This architecture, with its locally interacting elements and global update signals, can be formally identified as a cellular automata model (CA)\cite{tof}. As remarked earlier, CAs have been extensively studied mathematical abstractions, and it is well known that they can {\em in principle} support computation. In the majority of previous work\cite{nanotechArt} on CA-like architectures, each cell must be quite complex (possessing some internal processing capacity).  Here I have found that a CA can emerge directly from the physics of optically excited 2D nanoarrays, with each elementary quantum dot (or molecule) constituting a cell. Crucially I have established that the arrays can possess defects and yet be made perfectly functional {\em without knowing the nature or location of the defects}, and  this can be achieved while interfacing only at the edge. 

The architecture is suitable for self-assembly because it combines these properties with a periodic array pattern. By contrast, many previous novel computing schemes require a specific non-periodic pattern of cells in order to directly embed the algorithm. Such `hardwiring' would also exclude our evolutionary approach to defect tolerance. See e.g. the schemes in Refs. \onlinecite{benj2DCA} \& \onlinecite{smithRevsLent}. The latter is further limited by the need for very low temperatures.

Within the model described here, a fundamental limit on the rate at which cells can be switched is set by the decay rate from the unstable state $\ket{T}$ to the bistable states $\ket{0}$ \& $\ket{1}$. Consequently, since most optical transitions of this type tend to be in the nanosecond\cite{evenFastDecay} regime, it seems that the clock speed of a nanoarray device may not be superior to today's electronics. Indeed, once one allows for long pulses to ensure high switching reliability, the device may operate in MHz rather than GHz. However one should note that high clock speed is not the only route to computational power - this is demonstrated by the human brain, a massively parallel architecture operating with a switching rate of about 200 Hz. The nanoarray architecture advocated here would also benefit from massive parallelism, made possible by the coincidence of several factors: the small physical scale of cells and the low power dissipation per cell, the fact that optical dissipation avoids heating, the negligible per-cell fabrication cost due to self assembly, defect tolerance and wireless global control. Moreover, the architecture I have described does allow a high proportion of the cells to be usefully active at any given time. This is because the `processing' pulse sequence defined above (Fig. \ref{figFullAdder}) drives data through the programmed array {\em without} any net disruption of that program; therefore there can be a second independent set of data one period back in the array, and another behind that, and so on. This is an ultra-dense form of the {\em pipelining} strategy employed in modern CPUs. 

I stress that periodic array structures of the kind I envisage are already being experimentally realized. I hope this article will help to motivate work to study and engineer their internal states. The potential rewards are very great. By comparison to today's technology, these structures would be far smaller (a nanoarray of the size shown in Fig. \ref{figFullAdder} can occupy less area than a single transistor\cite{ITRS, Jian}), far cheaper to fabricate and far more energy efficient (a cell switching dissipation of $1$eV, five orders of magnitude below transistor technology, would be sufficient for room temperature stability). The architecture is wireless, with zero current flowing; indeed the flow of power into and out of the structure can be photonic, avoiding the thermal problems seen in today's CPUs. Consequently the architecture might be extended to 3D without any in-principle obstacles. One can envisage these structures acting as the computational core of future devices, with other complimentary nanostructures employed for data storage and bus functions.

 This work was supported by a Royal Society URF. I thank the Oxford Supercomputing Centre for donating CPU time.
 
\bigskip

\bigskip

\noindent {\Large \bf Appendix}

\bigskip
\noindent {\bf Pulse Sequences \& the Use of Cell `Types'}

The main paper states that a {\em sequence} of pulses can be used to systematically drive binary states though the array. Here I specify the pulse sequences, and elaborate on the roles of different cell `types'. I will use a notation where the symbol $\stackrel{ \rightarrow Z}{X_{Y}}$ denotes an optical pulse with frequency centered at $\omega^X_{2,1-Z}+(-1)^Z YK$. The quantities $\omega^X_{2,1-Z}$ and $K$ have been introduced in the main paper: $\omega^X_{2,1-Z}$ is the frequency to drive an {\em isolated} cell of type X from state $\ket{1-Z}$ into state $\ket{Z}$ via the transient state $\ket{T}$; $K$ is the characteristic strength of the cell-cell interaction between a pair of nearest neighbours (the strength elsewhere depends on the interaction form, c.f. Fig \ref{figComprehensive}b). The effect of such a pulse is to cause cells of the specified type to switch their state {\em conditional} on the perturbative effect of the neighbouring cells. Suppose I have an ideal array, one that is defect free and has a strict nearest neighbour interaction ($f({\bf r})=1$ if $|{\bf r}|\le$ cell separation, $=0$ otherwise). Then it is straightforward to determine the switching condition associated with a $\stackrel{ \rightarrow Z}{X_{Y}}$ pulse: the number of neighbouring cells in state $\ket{1}$ {\em minus} the number in state $\ket{0}$ must equal $Y$. So for example a pulse $\stackrel{\rightarrow 0}{A_{2}}$ will switch any $A$ cell currently in state $\ket{1}$ into $\ket{0}$ if there are exactly two more neighbours in state $\ket{1}$ than in state $\ket{0}$. Given the particular hexagonal geometry used in this paper (see Fig. \ref{figConcept}a), the
only way that this can happen is if the cell has a total of two neighbours, and both are in state $\ket{1}$. As a second example: a pulse $\stackrel{\rightarrow 1}{B_{-1}}$ will switch any $B$ cell currently in state $\ket{0}$ into $\ket{1}$ if it has exactly one less neighbour in state $\ket{1}$ than in state $\ket{0}$: this implies that the cell has a total of three neighbours of which two are in state $\ket{0}$. Notice that any cell with two neighbours can only possibly be switched by a pulse where $Y$ is $2$, $0$ or $-2$; by contrast any cell with three neighbours requires $Y$ from the set $\{-3,-1,1,3\}$. Thus a given pulse can only effect one of these two classes. This observation, combined with the particular arrangement of cell types $A$ and $B$, means that the geometry shown in Fig. \ref{figConcept}b does indeed meet condition (3) mentioned in the main paper: if a given cell is eligible to be switched by a certain pulse, then all the near neighbours will be ineligible, and therefore the neighbourhood will be stable for the duration of the pulse. This ensures that each pulse will drive the array into a single `attractor' even though the short time-scale dynamics are randomly asyncronous (see Fig. \ref{figDivergence}a).  If this condition were not respected, then a cell's neighbourhood, and hence its eligibility for switching, could change {\em during} the switching process. This could lead to non-deterministic  dynamics, as shown in Fig \ref{figDivergence}b.

\begin{figure*}[!t]
\centering
\resizebox{18.cm}{!}{\includegraphics{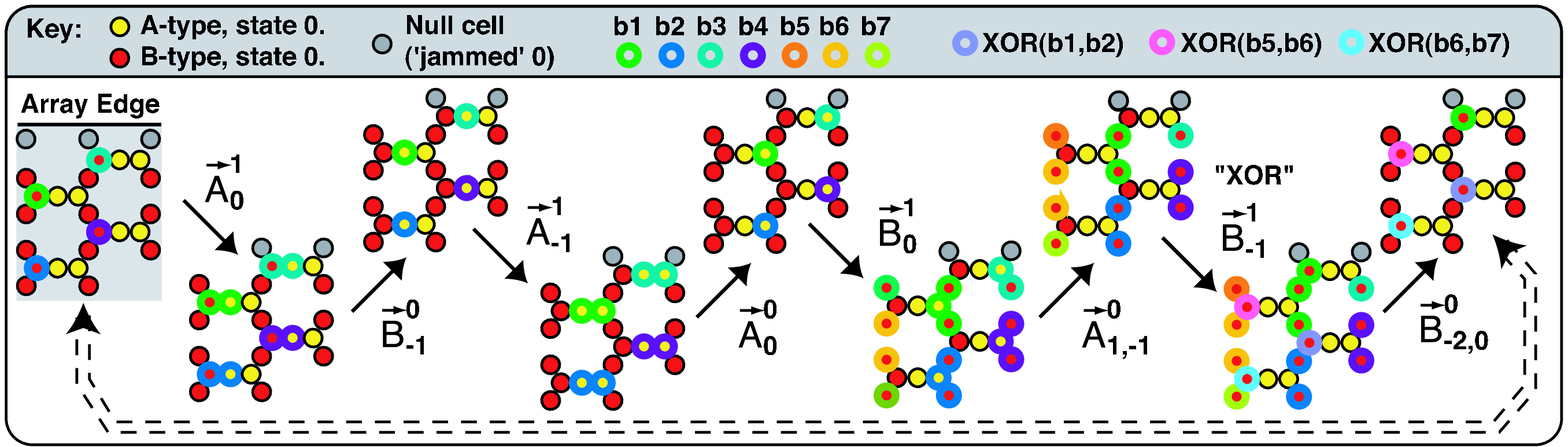}}
\caption{The {\em programming} pulse sequence in an ideal system: the figure shows how the cell states within a single fragment of the array develop as the sequence of global pulses is applied. The fragment is understood to be part of a regular array which extends away in three directions in the plane; the fourth in-plane direction (up) is the array edge, bounded by ``null cells'' that are permanently stuck in state $\ket{0}$. Filled yellow and red circles denote cells of type $A$ and $B$, respectively. Black bordered circles denote cells in state $\ket{0}$, colored borders indicate that a cell maybe in either state $\ket{0}$ or $\ket{1}$, and this state represents one {\em bit} of information. The sequence causes all bits to advance through one complete horizontal period of the array: each bit passes through $6$ cells including a fan-out node and an XOR node. Here $\stackrel{ \rightarrow 0}{A_1}_{,-1}$ denotes two pulses, $\stackrel{ \rightarrow 0}{A_{1}}$ and $\stackrel{ \rightarrow 0}{A_{-1}}$, applied in either order.}
\label{figProgramming}
\end{figure*}

 \begin{figure*}[!t]
  \begin{center}
    \leavevmode
\resizebox{18cm}{!}{\includegraphics{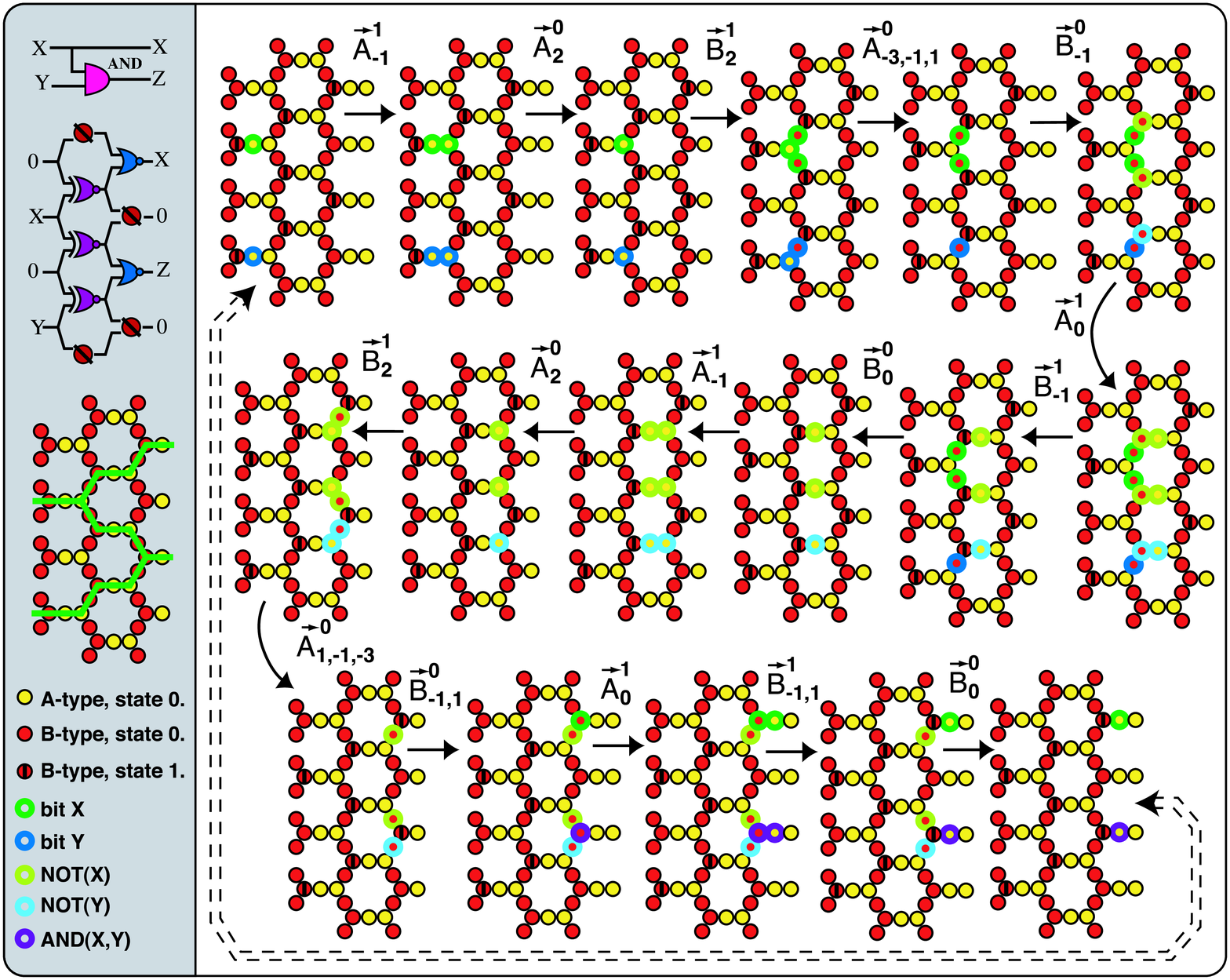}}
  \end{center}
\caption{The {\em processing} pulse sequence in an ideal array: the figure shows how the cell states within a single fragment of the array develop as the sequence is applied. Symbols are defined analogously to Fig. \ref{figProgramming}. By contrast to Fig. \ref{figProgramming}, here the initial state of the array has cells in state $\ket{1}$ at key points. These points effectively define the path of data through the array, allowing specific circuits to be defined even though the array is regular -- here the layout constitutes an AND gate as shown in the left panel. The pulse sequence causes data bits to advance through {\em two} horizontal periods of the array: each bit passes through $12$ cells including a fan-out node, a NXOR node, another fan-out node, and finally a NOR node. The sequence would then be repeated. Notice that the complete sequence does not disrupt the original the arrangement of $\ket{1}$ states. Hence pipelining is possible: in this geometry the maximum data density along the data path is one independent bit per eight cells.}
\label{figProcessing}
\end{figure*}

Given these properties, there may exist sequences of pulses that have the effect of systematically driving cell states across the array, effectively hopping from one cell to another. For an ideal system, such sequences may be discovered by a `pencil and paper' approach. Whether or not the intended behavior occurs in arrays with realistic interactions and multiple defects, is of course a crucial question one can subsequently tackle by numerical simulation. 

If we wish to be able to regard cell states as {\em bits} of information, to be processed by the internal dynamics of the array, then any pulse sequence must be `data blind' - it must be the same regardless of the specific states one wishes to drive. Two distinct variants must be found. The first is a {\em programming} sequence which propagates states into an initially empty array - a suitable sequence is shown in Fig. \ref{figProgramming}. After each repetition of the full sequence, the input cells would be set to their new values. Input cells are those on the far left of the array (not shown in Fig. \ref{figProgramming}, but see Fig \ref{figFullAdder}a) which represent the interface to other technologies, and are assumed to be directly controllable. This control of the left edge states, together with the pulse sequence, allows one to configure the internal array states. We then require a second {\em processing} pulse sequence whose effect is to drive data across the array, but only along paths defined by the pre-existing configuration of the array. This should be achieved without disrupting that configuration. A suitable sequence is defined in Fig. \ref{figProcessing}. Using these sequences one can a program the array for a specific function (such as binary addition), and process an arbitrary number of data sets. Multiple sets of input data could potentially be processed simultaneously: a kind of ultra-dense form of the `pipelining' technique used in conventional microprocessors. Then the array can be programmed for another function, etc. 

Having found pulse sequences for the idealized array, one can then move to numerical simulations to investigate whether functional behavior can also be developed for realistic arrays. The same pulse sequence is used, however each pulse is given a finite frequency width $f_w$, and additionally a small shift $\pm\Delta$ to allow for the fact that, at any given time, most non-nearest neighbours will be in state $\ket{0}$. The values $f_w$ and $\Delta$ are chosen manually and applied systematically to all pulses: a more refined approach would be to assign these values uniquely to each pulse and allow them to be optimized during the evolutionary process. This could significantly improve the tolerance of misalignment defects.

\bigskip
\noindent {\bf Fitness Criteria}

As depicted in Fig. \ref{figGenetic} of the main paper, the process of determining the fitness of a given genome involves a series of trials with different input values $x$ (in fact each $x$ is tried several times to check for randomness). In this way the GA obtains a mapping of input-to-output without any knowledge of the internal array dynamics. A fitness score is derived using the eight criteria listed below, each of which represents a different analysis of the input-output mapping. I use the following terminology: {\em input cells} are the cells on the far left of the array, {\em output cells} are those on the far right (as denoted by triangles in Fig \ref{figFullAdder}a). The set of bits placed onto the input cells is the {\em input state} $x$. The set of bits subsequently read from the output cells is the {\em output state}. An {\em interface point} is an output cell where processed data should exit the array according to the ideal target circuit. Thus Fig. \ref{figFullAdder}a has 7 output cells of which 2 are interface points; the intermediate stages shown in Fig \ref{figFullAdder}b have 3 interface points.

Note that some of the criteria can usefully differentiate between the poorly performing genomes that occur early in the evolutionary process, whereas other criteria become meaningful as the genome population approaches the ideal. The relative weights of the criteria were chosen manually and kept constant for all numerical experiments. This is an aspect that could be improved in future work: with sufficient numerical resources one could find the optimal weighting, at least for a given set of array characteristics (defect severity, interaction range etc.).

\noindent {\bf (1) Penalty for randomness}: negative score assigned if output state varies over repeated runs with a given input state $x$.  Penalty is super-linear in terms of the number of output cells at which randomness occurs, and is heavily enhanced if randomness occurs at an interface point.

\noindent {\bf (2) Connection to interface points}: Rewards when any interface point varies in {\em any} way as $x$ is varied.

\noindent {\bf (3) Connection across array from any input cell.} Investigates all cases where two input states $x$ differed at only one cell, and rewards when the corresponding two output states differ in {\em any} way. Reward is reduced if relevant randomness was found in criterion (1).  

\noindent {\bf (4) Correct input-output connections.} Similar to (3) but more strict: rewards only if the variation is at an interface point, and one that `should' be connected to the input cell in question according to the ideal target circuit.

\noindent {\bf (5) Bandwidth of connection}. (This criterion is only applied when randomness measured in (1) is below a certain threshold.) Evaluation of how many connections across the array can be inferred from the set of output states - e.g. the existence of 5 or more distinct output states implies a bandwidth of at least 3 bits. Higher bandwidths are rewarded, up to the ideal defined by the target circuit.

 \noindent {\bf (6) Crude matching to desired function}. For each instance where two output states, corresponding to two different $x$, differ from one another in any way, give a reward if and only if the target function calls for those $x$ to be differentiated. For example, in the half-adder circuit the inputs (A=1, B=0) and (A=0, B=1) should both generate the same output - therefore any variation in the measured output states from these inputs would not be rewarded by this criterion. 
 
 \noindent {\bf (7) Bitwise matching to desired function}. For each instance where two output states, corresponding to two different $x$, differ from one another {\em at an interface point}, give a reward if-and-only if  the target function calls for those $x$ to be differentiated at that interface point. For example, in the half-adder circuit the inputs (A=0, B=0) and (A=0, B=1) should give the same `carry bit' of $0$, but a different `sum bit'. Therefore variation at the interface point corresponding to the sum bit would be rewarded, whereas variation at the carry bit would not.
  
\noindent {\bf (8) Exact matching to desired function} The `ultimate' criterion that simply rewards when the states measured at an interface point are exactly as dictated by the target function. This criterion must reach maximum score for the circuit to have been achieved. If, in such a case, there is no randomness at the interface points, then the evolutionary process has been successfully concluded. 

\noindent {\bf (8)(weaker) Matching desired function to within a NOT.} A weaker version of criterion (8) was sometimes used for the intermediate stages of the evolutionary process: instead of demanding exact matching to the target function, it is permitted that interface points yield the inverse (i.e. the NOT) of desired value, providing that this is done consistently for all input states.


\bigskip 
\noindent {\bf Comments on the `Evolutionary Path'} 

Figure \ref{figFullAdder}b in the main paper illustrates the way in which I {\em impose} an evolutionary path by the use of intermediate goals. Instead of trying to directly produce the ideal complete circuit, I use the tactic of a set of targets of increasing complexity. Once a given goal is met, the GA switches immediately to the next. In this way one can minimize the chances of getting `stuck' in an evolutionary dead-end: a circuit that happens to score reasonably well by the 8 fitness criteria, but which cannot progress towards the ideal without significant `backtracking'. I have used a linear path, i.e. each evolutionary stage recognizes only one target. A more sophisticated approach, suited to more complex circuits, might define multiple possible paths. Applying such an approach to nanoarray simulations is an avenue for further work.

The tactic of using intermediate goals also provides a limited control over the spatial layout of the circuit. Notice that the second of the four targets in Fig. \ref{figFullAdder}b shows a half-adder circuit localized on the left side of the array. I can ensure that the circuit does indeed develop in that region by only permitting mutations in certain sections, or `genes', within the genomes (c.f. Fig \ref{figGenetic}). Specifically, I forbid mutations to genes lying in the first third of a genome. Thus since each gene is loaded into the array in order during the `programming' phase, I am forbidding mutations to the right-hand side of the circuit, even without understanding anything about the way the circuit is implemented by the genome~\cite{mightObject}. Once the evolutionary target is met, I move to a new target (e.g. the third circuit in Fig \ref{figFullAdder}b) and I may choose to exempt a different set of genes from the mutation process.

\bigskip 
\noindent{\bf Long range data transport}

An architecture of the kind outlined here is well suited to {\em processing} data: i.e. performing manipulations such as binary addition. However, it is not well suited to {\em transporting} data over long distances, because bits cannot move faster than one cell  per optical pulse. Therefore, one could envisage a functional device with a nanoarray processor at the heart, performing analogously to today's CPUs, while some other technology is used to interface this processor to data storage systems etc. Such an interface would presumably be `wire-like' rather than `array-like', and it is entirely plausible that this could be achieved at a molecular scale compatible with the array itself (given that various long conducting molecules have been observed, for example).

\bigskip 
%

\bigskip
\noindent{\bf Interface}

I have not discussed the details of the interface technology, i.e. the process by which cell states are set and read from the sides of the array. Those details would depend on specifics of the cell realization, beyond the 3-state abstraction that I have employed. But to give an example: if the cells along the array edge have unique transition energies (`type') then input can be achieved optically, and even optical readout is possible if we have a fourth state that decays to only one of the bistable states (thus making conditional fluorescence possible). Alternatively, if cell transition energies can be Stark shifted, then a set of lithographically defined electrodes along the array edge could be used to effectively tune the `type' of adjacent cells. 

Optical I/O schemes would operate at room temperature; if we are prepared to restrict ourselves to low temperatures, at least for preliminary experiments, then we can also consider the use of Coulomb blockade electrometers for readout. 

In a mature form of the technology the interface would presumably be to other molecular-scale elements.

\bigskip 

\noindent{\bf Quantum Information Processing}

All the analysis presented here is directed toward exploiting nanoarrays for {\em classical} information processing. Although the nanoarrays themselves are quantum structures, here we are using them to process {\em bits} rather than {\em qubits}, the fundamental units for quantum information processing (QIP). The underlying switching mechanism which I have employed is a dissipative, irreversible optical process, which makes the computation robust at room-temperature and gives it a clocked forward direction while avoiding the need for precisely timed laser pulses. This is ideal for our purposes, however the technique is completely unsuitable for manipulating qubits. Nevertheless it is possible that some of the ideas presented here could instead be applied to a coherent switching process. It is well established QC is formally possible in globally controlled systems, even in one-dimension~\cite{lloyd}. In our case we would look for a system where the transition $\ket{0} \leftrightarrow \ket{1}$ can be driven via state $\ket{T}$ in a coherent manner, by subjecting the system to two optical excitations simultaneously (i.e. a Raman transition, possibly a STIRAP process with overlapping short pulses). This is relatively straightforward for an isolated cell, although it requires precise pulse durations. {\em In principle} one could envisage a cell interacting with a network in such a way that applying a suitable two-mode pulse will produce differing {\em well defined} effects depending on the influence of its neighbours. In practice this might require a near perfect array: in particular the ubiquitous `misalignment defects', which we have seen successfully tolerated for classical processing, could be very damaging for QIP. However I note that near-perfect arrays {\em are} a possibility: for example, the structures described in the Li {\em et al} reference are apparently atomically precise in both cell composition and array geometry. In exploiting such an array, one might seek to engineer cells with unpaired electron spins to provide the low lying computational basis states, since they generally have superior decoherence properties. There is a substantial body of theoretical work aimed at QIP using spins in solid state structures, and optical control mechanisms have been explored~\cite{nanotechLoss}, e.g. through conditional creation of excitons. It should be noted that recently the cluster state formalism has broadened the range of physical systems that can perform QIP -- solid state arrays can support cluster state QIP~\cite{losSScluster}, but arrays without local addressing, as discussed here, may not. Moreover, now radically different non-array approaches, e.g. distributed matter qubits, may be competitive~\cite{LimPRA} as solid state QIP solutions. Finally, a relatively minor observation is that certain array geometries, including the hexagonal pattern I have employed here, are not well suited to the reversible ``2-into-2'' logic gates required for QIP. In summary, one cannot rule out the possibility of QIP in nanoarrays, but before such a goal could be considered realistic, one would first wish to demonstrate classical information processing of the kind described here.

\bigskip
\noindent{\bf Extensions}

This work is a first proof of principle and the numbers should therefore be understood as lower bounds on performance. The evolutionary process is far from optimal (several parameters were fixed arbitrarily), and this is an area for further work. With sufficient numerical resources one could also study larger arrays, realize larger and more diverse functions, and establish clearer trends. One might also extend the approach to quantum information processing.



\begin{thebibliography}{99}
\bibitem{fn1} I am assuming that the laser's spectral width is small compared to $E_0-E_1$.
\bibitem{fn2} Many physical systems have an inherent diagonal (Ising-type) interaction, and moreover off-diagonal components effectively vanish when the interacting systems are far off resonance. Thus this is a natural form of interaction to assume for systems with multiple cell `types'. Nevertheless not all arrays will be well modeled by this interaction: those which exhibit strong hopping of excitations will not be appropriate.
\bibitem{whyCon} This geometry can be realized with either two (Fig \ref{figConcept}a) or four (Fig. \ref{figComprehensive}b) cell types, thus allowing the performance in these two cases to be quantitatively compared.
\bibitem{evenFastDecay}Even if one were to find a system with a much faster decay, there is another independent constraint on the clock speed. In order to provide good frequency selectivity, one must use pulses that are long compared to the $\hbar/K$, since K gives the order of the detuning between cells that are `eligible' to be switched and those that are not.
\bibitem{mightObject}One might object that I am depending on the fact that genes loaded earlier during the `programming' procedure will propagate further to the right, and that when the GA makes use of this observations it is not strictly regarding the array as a `black-box'. However, the propagation order is practically a logical necessity given the way that wet are driving the array dynamics by laser pulse sequences - any array that is so `defective' as to violate this principle would be unusable.
\end{thebibliography}
\end{document}